\documentstyle[preprint,aps]{revtex}

\begin{document}

\draft
\title
{The Hausdorff dimension of fractal sets 
and fractional quantum Hall effect }

\author
{Wellington da Cruz}

\address
{Departamento de F\'{\i}sica,\\
 Universidade Estadual de Londrina, Caixa Postal 6001,\\
Cep 86051-970 Londrina, PR, Brazil\\}
 
\date{\today}

\maketitle

\begin{abstract}

We consider Farey series of rational numbers in terms of 
{\it fractal sets} labeled by 
the Hausdorff dimension with values defined 
in the interval $1$$\;$$ < $$\;$$h$$\;$$ <$$\;$$ 2$ and 
associated with fractal curves. Our 
results come from the observation that the fractional 
quantum Hall effect-FQHE occurs in pairs of {\it dual 
topological quantum numbers}, 
the filling factors. These quantum numbers obey some 
properties of the Farey series and so we obtain that {\it the 
universality classes of the quantum Hall transitions 
are classified in terms of $h$}. 
The connection between 
Number Theory and Physics appears naturally in this context.

\end{abstract}

\pacs{PACS numbers: 47.53.+n, 02.10.De; 05.30.-d; 05.70.Ce; 11.25.Hf 
\\
\\
E-mail address: wdacruz@exatas.uel.br}

\section{Introduction}

We have obtained from physical considerations about fractional quantum 
Hall efffect, a mathematical result related to the Farey 
series of rational numbers\cite{R1}. According to our approach the FQHE occurs in pairs 
of {\it dual topological quantum numbers}, the filling factors. These parameters 
characterize the quantization of the Hall resistance in some systems of the 
condensed matter under lower temperatures and intense external magnetic fields. 
The filling factor is defined by 
$f=N\frac{\phi_{0}}{\phi}$, where $N$ is the electron number, 
$\phi_{0}$ is the quantum unit of flux and
$\phi$ is the flux of the external magnetic field throughout the sample. In our 
formulation {\it the filling factor gets its topological character from 
the parameter $h$} to be defined.

We can check that {\it all 
experimental data for the occurrence of FQHE} satisfies a 
{\it symmetry principle} discovered by us, that is, the {\it duality 
symmetry between universal classes $h$ of particles or quasiparticles} 
with any value of spin\cite{R1,R2,R3,R4,R5}. For example, 
we have the dual filling factors
\\

$(f,\tilde{f})=\left(\frac{1}{3},\frac{2}{3}\right), 
\left(\frac{5}{3},\frac{4}{3}\right), \left(\frac{1}{5},\frac{4}{5}\right), 
\left(\frac{2}{7},\frac{5}{7}\right),\left(\frac{2}{9},\frac{7}{9}\right), 
\left(\frac{2}{5},\frac{3}{5}\right), \left(\frac{3}{7},\frac{4}{7}\right), 
\left(\frac{4}{9},\frac{5}{9}\right) etc$.\\

On the other hand, in Ref.\cite{R1} we have considered the 
filling factors into equivalence classes 
labeled by $h$ and so, we can estimate the occurrence 
of FQHE, just taking into account a fractal spectrum and 
a duality symmetry between the classes. A relation 
between equivalence classes $h$ and the modular group\cite{R6,R1} for 
the quantum phase transitions of the FQHE was also noted. As a consequence, 
the parameter {\it $h$ classifies 
the universality class of these transitions}. Another approach 
in the literature\cite{R7} has considered Cantor sets in 
connection with the FQHE and some relation with the Farey sequences.

\section{Hausdorff dimension, fractal sets and Farey series}

Now, considering fractal curves in connection with quantum paths, we have defined 
a {\it fractal spectrum}, which relates the parameter $h$ and the spin $s$ 
of the particle through the spin-statistics relation defined by 
$\nu=2s=2\frac{\phi\prime}{\phi_{0}}$, where 
$\phi\prime$ is the magnetic flux 
associated to the charge-flux system. Follows that, 

\begin{eqnarray}
&&h-1=1-\nu,\;\;\;\; 0 < \nu < 1;\;\;\;\;\;\;\;\;
 h-1=\nu-1,\;
\;\;\;\;\;\; 1 <\nu < 2;\\
&&h-1=3-\nu,\;\;\;\; 2 < \nu < 3;\;\;\;\;\;\;\;\;
 h-1=\nu-3,\;
\;\;\;\;\;\; 3 <\nu < 4;\\
&&h-1=5-\nu,\;\;\;\; 4 < \nu < 5;\;\;\;\;\;\;\;\;
 h-1=\nu-5,\;
\;\;\;\;\;\; 5 <\nu < 6;\\
&&h-1=7-\nu,\;\;\;\; 6 < \nu < 7;\;\;\;\;\;\;\;\;
 h-1=\nu-7,\;
\;\;\;\;\;\; 7 <\nu < 8;\\
&&etc.\nonumber
\end{eqnarray}

\noindent where $h$ is a fractal parameter or Hausdorff 
dimension defined into the interval 
$1$$\;$$ < $$\;$$h$$\;$$ <$$\;$$ 2$. According to our approach 
there is a correspondence between $f$ and $\nu$, 
numerically $f=\nu$.

The fractal curve is continuous 
and nowhere differentiable, it is self-similar, it does not depend 
on the scale and has fractal dimension just in that interval. Given a closed path 
with length $L$ and resolution $R$, the fractal properties of this curve 
can be determined by 

\begin{eqnarray}
h-1=\lim_{R\rightarrow 0}\frac{\ln{L/l}}{\ln R},
\end{eqnarray}

\noindent where $l$ is the usual length for the resolution $R$ and 
the curve is covering with $l/R$ spheres of diameter $R$.

Farey series $F_{n}$ of order $n$ is the increasing sequence of 
irreducible fractions in the range $ 0 $ to $ 1$ whose 
denominators do not exceed $n$. They satisfy the properties

P1. If $\nu_{1}=\frac{p_{1}}{q_{1}}$ and 
$\nu_{2}=\frac{p_{2}}{q_{2}}$ are two consecutive fractions 
$\frac{p_{1}}{q_{1}}$$ >$$ \frac{p_{2}}{q_{2}}$, then 
$|p_{2}q_{1}-q_{2}p_{1}|=1$.

P2. If $\frac{p_{1}}{q_{1}}$, $\frac{p_{2}}{q_{2}}$,
$\frac{p_{3}}{q_{3}}$ are three consecutive fractions 
$\frac{p_{1}}{q_{1}}$$ >$$ \frac{p_{2}}{q_{2}} 
$$>$$ \frac{p_{3}}{q_{3}}$, then 
$\frac{p_{2}}{q_{2}}=\frac{p_{1}+p_{3}}{q_{1}+q_{3}}$.

P3. If $\frac{p_{1}}{q_{1}}$ and $\frac{p_{2}}{q_{2}}$ are 
consecutive fractions in the same sequence, then among 
all fractions\\
 between the two, 
$\frac{p_{1}+p_{2}}{q_{1}+q_{2}}$
 is the unique reduced
fraction with the smallest denominator.

Let us consider the {\it fractal spectrum} and the {\it duality 
symmetry} between the sets $h$, defined by ${\tilde{h}}=3-h$, so we have
the following

\newpage

{\bf Theorem}:

{\it The elements of the Farey series $F_{n}$ 
of the order $n$, belong to the fractal sets, whose Hausdorff 
dimensions are the second fractions of the fractal sets. The 
Hausdorff dimension has values within the interval 
$1$$\;$$ < $$\;$$h$$\;$$ <$$\;$$ 2$, which are associated with fractal curves.}

For example, consider the Farey series of order $6$ for any interval
\vspace{10mm}
\begin{center}
Table 
\end{center}
\begin{center}
\begin{tabular}{|c|c|c|c|c|c|c|c|c|c|c|c|c|c|}
\hline
$h$ & $2$ & $\frac{11}{6}$ & $\frac{9}{5}$ & $\frac{7}{4}$  & $\frac{5}{3}$
& $\frac{8}{5}$  & $\frac{3}{2}$ &  $\frac{7}{5}$ & $\frac{4}{3}$ & $\frac{5}{4}$ &
 $\frac{6}{5}$ & $\frac{7}{6}$ & $ 1$\\
\hline
$0<\nu<1$ & $\frac{0}{1}$ & $\frac{1}{6}$ & $\frac{1}{5}$ & $\frac{1}{4}$  
& $\frac{1}{3}$& $\frac{2}{5}$ & $\frac{1}{2}$ & $\frac{3}{5}$& 
$\frac{2}{3}$ & $\frac{3}{4}$ & $\frac{4}{5}$ & $\frac{5}{6}$ & $\frac{1}{1}$\\
\hline
$1<\nu<2$ & $\frac{2}{1}$ & $\frac{11}{6}$ & $\frac{9}{5}$ & $\frac{7}{4}$  
& $\frac{5}{3}$&
$\frac{8}{5}$ & $\frac{3}{2}$ & $\frac{7}{5}$ 
& $\frac{4}{3}$ & $\frac{5}{4}$ & $\frac{6}{5}$ & $\frac{7}{6}$ & $\frac{1}{1}$\\
\hline
$2<\nu<3$ & $\frac{2}{1}$ & $\frac{13}{6}$ & $\frac{11}{5}$ & $\frac{9}{4}$  
& $\frac{7}{3}$&
$\frac{12}{5}$ & $\frac{5}{2}$ & $\frac{13}{5}$ 
& $\frac{8}{3}$ & $\frac{11}{4}$ & $\frac{14}{5}$ & $\frac{17}{6}$ & $\frac{3}{1}$\\
\hline
$3<\nu<4$ & $\frac{4}{1}$ & $\frac{23}{6}$ & $\frac{19}{5}$ & $\frac{15}{4}$  
& $\frac{11}{3}$&
$\frac{18}{5}$ & $\frac{7}{2}$ & $\frac{17}{5}$ 
& $\frac{10}{3}$ & $\frac{15}{4}$ & $\frac{16}{5}$ & $\frac{19}{6}$ & $\frac{3}{1}$\\
\hline
$4<\nu<5$ & $\frac{4}{1}$& $\frac{25}{6}$ & $\frac{21}{5}$ & $\frac{17}{4}$  
& $\frac{13}{3}$&
$\frac{22}{5}$ & $\frac{9}{2}$ & $\frac{23}{5}$ 
& $\frac{14}{3}$ & $\frac{19}{4}$ & $\frac{24}{5}$ & $\frac{29}{6}$ & $\frac{5}{1}$\\
\hline
$5<\nu<6$ & $\frac{6}{1}$ & $\frac{35}{6}$ & $\frac{29}{5}$ & $\frac{23}{4}$  
& $\frac{17}{3}$&
$\frac{28}{5}$ & $\frac{11}{2}$ & $\frac{27}{5}$ 
& $\frac{16}{3}$ & $\frac{21}{4}$ & $\frac{26}{5}$ & $\frac{31}{6}$ & $\frac{5}{1}$\\
\hline
$6<\nu<7$ & $\frac{6}{1}$ & $\frac{37}{6}$ & $\frac{31}{5}$ & $\frac{25}{4}$  
& $\frac{19}{3}$&
$\frac{32}{5}$ & $\frac{13}{2}$ & $\frac{33}{5}$ 
& $\frac{20}{3}$ & $\frac{27}{4}$ & $\frac{34}{5}$ & $\frac{41}{6}$ & $\frac{7}{1}$\\
\hline
$7<\nu<8$ & $\frac{8}{1}$& $\frac{47}{6}$ & $\frac{39}{5}$ & $\frac{31}{4}$  
& $\frac{23}{3}$&
$\frac{38}{5}$ & $\frac{15}{2}$ & $\frac{37}{5}$ 
& $\frac{22}{3}$ & $\frac{29}{4}$ & $\frac{36}{5}$ & $\frac{43}{6}$ & $\frac{7}{1}$\\
\hline
$8<\nu<9$ & $\frac{8}{1}$& $\frac{49}{6}$ & $\frac{41}{5}$ & $\frac{33}{4}$  
& $\frac{25}{3}$&
$\frac{42}{5}$ & $\frac{17}{2}$ & $\frac{43}{5}$ 
& $\frac{26}{3}$ & $\frac{35}{4}$ & $\frac{44}{5}$ & $\frac{53}{6}$ & $\frac{9}{1}$\\
\hline

$9<\nu<10$ & $\frac{10}{1}$& $\frac{59}{6}$ & $\frac{49}{5}$ & $\frac{39}{4}$  
& $\frac{29}{3}$&
$\frac{48}{5}$ & $\frac{19}{2}$ & $\frac{47}{5}$ 
& $\frac{28}{3}$ & $\frac{37}{4}$ & $\frac{46}{5}$ & $\frac{55}{6}$ & $\frac{9}{1}$\\
\hline

$10<\nu<11$ & $\frac{10}{1}$& $\frac{61}{6}$ & $\frac{51}{5}$ & $\frac{41}{4}$  
& $\frac{31}{3}$&
$\frac{52}{5}$ & $\frac{21}{2}$ & $\frac{53}{5}$ 
& $\frac{32}{3}$ & $\frac{43}{4}$ & $\frac{54}{5}$ & $\frac{65}{6}$ & $\frac{11}{1}$\\
\hline

$11<\nu<12$ & $\frac{12}{1}$& $\frac{71}{6}$ & $\frac{59}{5}$ & $\frac{47}{4}$  
& $\frac{35}{3}$&
$\frac{58}{5}$ & $\frac{23}{2}$ & $\frac{57}{5}$ 
& $\frac{34}{3}$ & $\frac{45}{4}$ & $\frac{56}{5}$ & $\frac{67}{6}$ & $\frac{11}{1}$\\
\hline

$12<\nu<13$ & $\frac{12}{1}$& $\frac{73}{6}$ & $\frac{61}{5}$ & $\frac{49}{4}$  
& $\frac{37}{3}$&
$\frac{62}{5}$ & $\frac{25}{2}$ & $\frac{63}{5}$ 
& $\frac{38}{3}$ & $\frac{51}{4}$ & $\frac{64}{5}$ & $\frac{77}{6}$ & $\frac{13}{1}$\\
\hline

$13<\nu<14$ & $\frac{14}{1}$& $\frac{83}{6}$ & $\frac{69}{5}$ & $\frac{55}{4}$  
& $\frac{41}{3}$&
$\frac{68}{5}$ & $\frac{27}{2}$ & $\frac{67}{5}$ 
& $\frac{40}{3}$ & $\frac{53}{4}$ & $\frac{66}{5}$ & $\frac{79}{6}$ & $\frac{13}{1}$\\
\hline

$14<\nu<15$ & $\frac{14}{1}$& $\frac{85}{6}$ & $\frac{71}{5}$ & $\frac{57}{4}$  
& $\frac{43}{3}$&
$\frac{72}{5}$ & $\frac{29}{2}$ & $\frac{73}{5}$ 
& $\frac{44}{3}$ & $\frac{59}{4}$ & $\frac{74}{5}$ & $\frac{89}{6}$ & $\frac{15}{1}$\\
\hline

$15<\nu<16$ & $\frac{16}{1}$& $\frac{95}{6}$ & $\frac{79}{5}$ & $\frac{63}{4}$  
& $\frac{47}{3}$&
$\frac{78}{5}$ & $\frac{31}{2}$ & $\frac{77}{5}$ 
& $\frac{46}{3}$ & $\frac{61}{4}$ & $\frac{76}{5}$ & $\frac{91}{6}$ & $\frac{15}{1}$\\
\hline

$16<\nu<17$ & $\frac{16}{1}$& $\frac{97}{6}$ & $\frac{81}{5}$ & $\frac{65}{4}$  
& $\frac{49}{3}$&
$\frac{82}{5}$ & $\frac{33}{2}$ & $\frac{83}{5}$ 
& $\frac{50}{3}$ & $\frac{67}{4}$ & $\frac{84}{5}$ & $\frac{101}{6}$ & $\frac{17}{1}$\\
\hline

$17<\nu<18$ & $\frac{18}{1}$& $\frac{107}{6}$ & $\frac{89}{5}$ & $\frac{71}{4}$  
& $\frac{53}{3}$&
$\frac{88}{5}$ & $\frac{35}{2}$ & $\frac{87}{5}$ 
& $\frac{52}{3}$ & $\frac{69}{4}$ & $\frac{86}{5}$ & $\frac{103}{6}$ & $\frac{17}{1}$\\
\hline

$\cdots$ & $\cdots $ & $\cdots $ & $\cdots$ & $\cdots$  & $\cdots$&
$\cdots$ & $\cdots$ & $\cdots$ 
& $\cdots$ & $\cdots$ & $\cdots$ & $\cdots$ &$ \cdots$\\
\hline
\end{tabular}
\end{center}
\vspace{5mm}

\newpage

Then, we obtain {\it fractal sets} labeled by the Hausfdorff dimension
\\
\\

\begin{eqnarray}
&&\biggl\{\frac{1}{6},\frac{11}{6},\frac{13}{6},\frac{23}{6},\frac{25}{6},\frac{35}{6},
\frac{37}{6},\frac{47}{6},\frac{49}{6},\frac{59}{6},\frac{61}{6},
\cdots\biggr\}_{h=\frac{11}{6}};\nonumber\\
&&\biggl\{\frac{1}{5},\frac{9}{5},\frac{11}{5},\frac{19}{5},\frac{21}{5},\frac{29}{5},
\frac{31}{5},\frac{39}{5},\frac{41}{5},\frac{49}{5},\frac{51}{5},
\cdots\biggr\}_{h=\frac{9}{5}};\nonumber\\
&&\biggl\{\frac{1}{4},\frac{7}{4},\frac{9}{4},\frac{15}{4},\frac{17}{4},\frac{23}{4},
\frac{25}{4},\frac{31}{4},\frac{33}{4},\frac{39}{4},\frac{41}{4},
\cdots\biggr\}_{h=\frac{7}{4}};\nonumber\\
&&\biggl\{\frac{1}{3},\frac{5}{3},\frac{7}{3},\frac{11}{3},\frac{13}{3},\frac{17}{3},
\frac{19}{3},\frac{23}{3},\frac{25}{3},\frac{29}{3},\frac{31}{3},
\cdots\biggr\}_{h=\frac{5}{3}}\nonumber\\
&&\biggl\{\frac{2}{5},\frac{8}{5},\frac{12}{5},\frac{18}{5},\frac{22}{5},\frac{28}{5},
\frac{32}{5},\frac{38}{5},\frac{42}{5},\frac{48}{5},\frac{52}{5},
\cdots\biggr\}_{h=\frac{8}{5}};\nonumber\\
&&\biggl\{\frac{1}{2},\frac{3}{2},\frac{5}{2},\frac{7}{2},\frac{9}{2},\frac{11}{2},
\frac{13}{2},\frac{15}{2},\frac{17}{2},\frac{19}{2},\frac{21}{2},
\cdots\biggr\}_{h=\frac{3}{2}};\\
&&\biggl\{\frac{3}{5},\frac{7}{5},\frac{13}{5},\frac{17}{5},\frac{23}{5},\frac{27}{5},
\frac{33}{5},\frac{37}{5},\frac{43}{5},\frac{47}{5},\frac{53}{5},
\cdots\biggr\}_{h=\frac{7}{5}};\nonumber\\
&&\biggl\{\frac{2}{3},\frac{4}{3},\frac{8}{3},\frac{10}{3},\frac{14}{3},\frac{16}{3},
\frac{20}{3},\frac{22}{3},\frac{26}{3},\frac{28}{3},\frac{32}{3},
\cdots\biggr\}_{h=\frac{4}{3}};\nonumber\\
&&\biggl\{\frac{3}{4},\frac{5}{4},\frac{11}{4},\frac{13}{4},\frac{19}{4},\frac{21}{4},
\frac{27}{4},\frac{29}{4},\frac{35}{4},\frac{37}{4},\frac{43}{4},
\cdots\biggr\}_{h=\frac{5}{4}};\nonumber\\
&&\biggl\{\frac{4}{5},\frac{6}{5},\frac{14}{5},\frac{16}{5},\frac{24}{5},\frac{26}{5},
\frac{34}{5},\frac{36}{5},\frac{44}{5},\frac{46}{5},\frac{54}{5},
\cdots\biggr\}_{h=\frac{6}{5}};\nonumber\\
&&\biggl\{\frac{5}{6},\frac{7}{6},\frac{17}{6},\frac{19}{6},\frac{29}{6},\frac{31}{6},
\frac{41}{6},\frac{43}{6},\frac{53}{6},\frac{55}{6},\frac{65}{6},
\cdots\biggr\}_{h=\frac{7}{6}}.\nonumber
\end{eqnarray}
\\
\\
Observe that the sets are dual sets and, in particular, we have 
a fractal selfdual set, with Hausdorff dimension $h=\frac{3}{2}$. Thus, in this 
way we can extract for any Farey series of rational numbers, {\it taking into account the 
fractal spectrum and the duality symmetry between sets}, fractal 
sets whose Hausdorff dimension is the second fraction of the set.

By another method, we have obtained for {\it fractons} or charge-flux systems, that is, 
particles with any value of spin defined 
in two-dimensional multiply connected space, a {\it fractal distribution function}\cite{R1,R2} 

\begin{eqnarray}
\label{e.h} 
n=\frac{1}{{\cal{Y}}[\xi]-h}
\end{eqnarray}

\noindent where ${\cal{Y}}[\xi]$ satisfies the equation  

\begin{eqnarray}
\label{e.1} 
\xi=\biggl\{{\cal{Y}}[\xi]-1\biggr\}^{h-1}
\biggl\{{\cal{Y}}[\xi]-2\biggr\}^{2-h}
\end{eqnarray}

\noindent and $\xi=\exp\left\{(\epsilon-\mu)/KT\right\}$ has the usual definition.

This {\it quantum-geometrical} description of the statistical laws of nature 
is associated with a {\it fractal von Neumann entropy } per state in terms of 
the average occupation number

\begin{eqnarray}
{\cal{S}}_{G}[h,n]=K\left[\left[1+(h-1)n\right]\ln\left\{\frac{1+(h-1)n}{n}\right\}
-\left[1+(h-2)n\right]\ln\left\{\frac{1+(h-2)n}{n}\right\}\right].
\end{eqnarray}

An interesting point is that the solutions for the algebraic 
equations given by the Eq.(\ref{e.1}) are of the form

\[ 
{\cal{Y}}_{h}[\xi]=f[\xi]+{\tilde{h}}
\]

or

\[ 
{\cal{Y}}_{\tilde{h}}[\xi]=g[\xi]+h.
\]

The functions $f[\xi]$  and $g[\xi]$ at least 
for third, fourth degrees algebraic equation differ by signals $\pm$ 
in some terms of their expressions. Observe also that the solution for a given $h$ 
receives its dual $\tilde{h}$ as a constant. We can conjecture if this result gives us 
some information about these classes of algebraic equations. In Ref.\cite{R3} 
we have also defined a topological concept termed {\it fractal index} related to the 
Rogers dilogarithm function and the concept of central charge 
associated with the conformal field theories. The fractal index can 
assume rational values and are related to the algebraic numbers which come 
from of the algebraic equation for each value of $h$ by the Eq.(\ref{e.1}). 
The connection between Physics and Number Theory is manifest because, on the one hand, 
we have the FQHE characterized in terms of a geometrical parameter related 
to the fractal curves and, on the other hand, the dilogarithm function 
appears in various branches of mathematics besides number theory, 
such as\cite{R8}: exactly solvable 
models, algebraic K-theory, hyperbolic manifolds, low dimensional 
topology etc. Now, as was introduced 
in\cite{R3}, we have also established a connection between 
{\it fractal geometry and Rogers dilogarithm function}.

\section{Conclusions}

In this Letter {\it we have determined an algorithm for computation 
of the Hausdorff dimension of any 
fractal set related to the Farey series}. This mathematical result is 
related to the observation that {\it the FQHE occurs in pairs of 
dual topological filling factors}\cite{R1}. 
In this way, {\it these quantum numbers 
are classified in classes labeled by the parameter $h$ and satisfy the 
properties of the Farey sequences}. Thus, {\it the universality classes 
of the quantum Hall transitions are classified in terms of $h$} and 
this result is obtained from our analysis considering the modular 
group for these transitions\cite{R1,R6}. We have also, 
in our context, made {\it a connection between Physics and Number Theory relating the 
fractal geometry and dilogarithm function through of the concept of fractal 
index} introduced by us\cite{R3}. We have also the {\it possibility of fractal sets 
with irrational values for the Hausdorff dimension}.

\acknowledgments

The author thanks the referee by the comments and suggestions.

\end{document}